\documentstyle[12pt,axodraw,epsfig]{article}

\newcommand{\be}{\begin{equation}}
\newcommand{\bea}{\begin{eqnarray}}
\newcommand{\ba}{\begin{array}}
\newcommand{\bean}{\begin{eqnarray*}}
\newcommand{\ee}{\end{equation}}
\newcommand{\eea}{\end{eqnarray}}
\newcommand{\ea}{\end{array}}
\newcommand{\eean}{\end{eqnarray*}}

\def \Dsl {D \kern-.65em{/}}
\def \nonSM {SM \kern-.75em{/}} 
\def \qsl {q \kern-.45em{/}}
\def \slp {p \kern-.45em{/}}
\def \ksl {k \kern-.45em{/}}
\def \lsl {\ell \kern-.45em{/}}
\def \dsl {p_d \kern-.85em{/}}
\def \usl {p_u \kern-.90em{/}}
\def \dk {\Delta\kappa_{\gamma}}
\def \dk2 {\Delta^2\kappa_{\gamma}}

\def \d1 {\delta_{1}}
\def \d2 {\delta_{2}}
\def \d3 {\delta_{3}}
\def \d4 {\delta_{4}}

\begin{document}

\begin{flushright}
CERN-TH/99-212\\
July 1999
\end{flushright}
 
\begin{center}
{\Large {\bf
 Probing the $WW\gamma$ vertex 
at hadron colliders}}\\[2.4cm]
{\large Joannis Papavassiliou}$^a$ 
                           {\large and Kostas Philippides}$^b$\\[0.4cm]
$^a${\em Theory Division, CERN, CH-1211 Geneva 23, Switzerland}\\[0.3cm]
$^b${\em Department of Physics, University of Durham, Durham, DH1 3LE, 
U.K.}
\end{center}
 
\vskip0.7cm     \centerline{\bf   ABSTRACT}  \noindent

We present a  new, model independent method  for extracting bounds for
the anomalous $\gamma WW$ couplings from  hadron collider experiments. 
At the  partonic level we introduce  a set  of three observables which
are constructed  from  the unpolarized differential  cross-section for
the process  $d\bar{u}\to W^{-}\gamma$ by appropriate convolution with
a  set of simple   polynomials  depending only on  the  center-of-mass
angle.  One of these  observables allows for the  direct determination
of  the anomalous coupling usually  denoted by $\Delta\kappa$, without
any simplifying assumptions, and without  relying on the presence of a
radiation zero.  The other two observables impose two sum rules on the
remaining three anomalous  couplings.  The inclusion  of the structure
functions is    discussed  in detail for  both     $p\bar{p}$ and $pp$
colliders.  We show that,  whilst for $p\bar{p}$ experiments  this can
be accomplished straightforwardly, in the $pp$  case one has to resort
to somewhat  more elaborate techniques, such  as the binning of events
according to their longitudinal momenta.

\vskip0.7cm


\newpage

\setcounter{equation}{0}
\section{Introduction}

The importance of hadron colliders  concerning the measurement of  the
trilinear  gauge  self-couplings  has  already been exemplified by the
direct  verification   of  the  existence   of  such  vertices through
$W\gamma$ as  well as $WZ$ production  at  the Tevatron \cite{Hadron}. 
The bounds on the  anomalous couplings so  obtained have only recently
been surpassed by the LEP2 measurements \cite{LEP}.  The advent of the
Large  Hadron Collider  (LHC) at   CERN  is  expected to improve   the
situation further, due to  its high  luminosity  and energy  reach.  A
significant advantage of hadron colliders  in this context is the fact
that the couplings of the  photon and the $Z$  boson to the $W$  boson
can be probed independently through separate processes. 
Taking also into account that at the sub-process level 
 the center-of-mass energy varies, one may also 
study the form-factor structure of the 
 couplings, thus furnishing
important information, complementary to that 
 obtained from the lepton colliders.

In this paper we extend 
to hadron 
collider experiments
a model-independent method proposed 
for the extraction 
of bounds on the anomalous couplings 
at LEP2 \cite{sumrules}. 
This method is 
based on constructing appropriate projections of the 
differential cross section \cite{PRW}, which 
lead to a set of novel observables; the latter  
are related to the anomalous couplings by means 
of simple 
algebraic equations. 
The experimental determinations of these 
observables can in turn
be used in order to impose bounds  simultaneously on all 
anomalous couplings, without having to resort to 
model-dependent 
relations among them, or invoke any
further simplifying assumptions.

Given the advantages mentioned above,
the generalization of this method to the case of 
hadron colliders is certainly interesting, 
but is by no means
obvious, even at the theoretical level.
The crucial difference is that at hadron colliders
the center-of-mass energy of the sub-processes is
not fixed, and the effects of the 
structure functions must be included.
The detailed study reveals however that, due to
a very particular dependence of
the differential cross-section on the center-of-mass angle, 
the method can in fact be applied. 
As a result, one obtains a set of three algebraic equations 
relating the four unknown couplings. In particular, one 
can directly extract the value for the $\Delta \kappa$ anomalous
form-factor,
{\it without} having to assume the absence
of other anomalous couplings. 

The paper is organized as follows:
 In section 2 we present explicit results of the partonic 
 differential 
 cross-section  
for the prototype process
$d\bar{u}\to W^{\pm}\gamma$, using the most general 
  $WW\gamma$ vertex allowed from Lorentz and $U(1)_{EM}$ invariance. 
This process  
has received significant 
attention in the literature 
\cite{MSS},\cite{Robi},\cite{Hagi},\cite{BZ88},\cite{BB90}, mainly 
 due to the presence of the radiation zero in its differential 
cross-section \cite{radzero1},\cite{radzero2}.
In section 3 we define a set of observables, 
which depend explicitly
on the various anomalous couplings, and can be experimentally
extracted through
the convolution of the differential cross-section
with appropriately constructed polynomials of $\cos\theta$,
the center-of-mass 
scattering angle. In section 4 we present an elementary
discussion of the statistical properties of these observables.
In section 5 we turn our attention to  the 
realistic cases of $pp$ and  $p\bar{p}$ collisions, and address the 
complications that arise due to the inclusion of the structure functions.
Finally, in section 6 we present our conclusions.

\setcounter{equation}{0}
\section{The process $d\bar{u}\to W^{-}\gamma$ 
in the presence of anomalous couplings}

We consider the process 
$d(p_d)\bar{u}(p_u) \rightarrow W^{-}(p_2)\gamma(q)$,
shown in Fig.~1.
All momenta are incoming 
i.e.
$p_{d}+p_{u}+p_2+q=0$, and the square of center-of-mass energy,
$s$, is given by $s= p_1^2=(p_d+p_u)^2 =(p_2+q)^2$.
We work in the narrow width approximation where the $W$ is 
assumed  to be strictly on shell, $p_2^2=M_W^2$; 
the inclusion of $W$ off-shellness effects 
is known to give very small contributions \cite{Hagi}.

The $S$-matrix element for the process we study is given by
\be
\langle d \bar{u} |W^{-} \gamma \rangle =
T^{\mu\beta} \epsilon^{\mu}_{\gamma}(q)\epsilon^{\beta}_{W^-}(p_2)~,
\ee
where
$\epsilon^{\mu}_{\gamma}(q)$ and $\epsilon^{\beta}_{W^-}(p_2)$
are the polarization vectors of the photon and the $W$, respectively,
and the amplitude $T^{\mu\beta}$
is the sum of three pieces,
\be
T^{\mu\beta}=T_s^{\mu\beta}+T_t^{\mu\beta}+T_u^{\mu\beta}~,
\ee
where $T_s$, $T_t$ and $T_u$ denote 
the $s$, $t$ and $u$ channels contributions, respectively.
They are given 
by the following expressions: 
\bea
T_t^{\mu\beta}&=& \left(\frac{iegQ_{d}}{\sqrt{2}}\right) \ 
\bar{v}_u\gamma_{\beta} P_L\frac{1}{\dsl+\qsl}\gamma_{\mu} u_d~, 
\nonumber\\ 
T_u^{\mu\beta} &=& \left(\frac{iegQ_{\bar{u}} }{\sqrt{2}}\right)\ 
\bar{v}_u\gamma_{\mu}\frac{1}{\usl+\qsl}\gamma_{\beta} P_L u_d~,
\nonumber\\
T_s^{\mu\beta} &=& \left(\frac{-ieg}{\sqrt{2}}\right) \ 
J_{\alpha}^{W^{-}}
\left(\frac{1}{p_1^2-M_W^2}\right)
\Gamma^{\mu\alpha\beta}(q,p_1,p_2)~,
\eea 
where 
$J_{\alpha}^{W^{-}}=\bar{v}_u\gamma_{\alpha}P_Lu_d$
$P_L=\frac{1}{2}(1-\gamma_5)$, $Q_{\bar{u}}=2Q_{d}=-\frac{2}{3}$, and
we have neglected any quark mixing effects. 
$\Gamma^{\mu\alpha\beta}$ denotes the $WW\gamma$ vertex;
it is written as the sum of the usual Standard Model piece 
$\Gamma_{0}^{\mu\alpha\beta}$, 
and an anomalous piece $\Gamma^{\mu\alpha\beta}_{an}$, i.e.
\be
\Gamma^{\mu\alpha\beta}(q,p_1,p_2)=
\Gamma_{0}^{\mu\alpha\beta}(q,p_1,p_2)
+\Gamma^{\mu\alpha\beta}_{an}(q,p_1,p_2)~.
\ee
The canonical piece $\Gamma_{0}^{\mu\alpha\beta}$ is given by
\be
\Gamma_{0}^{\mu\alpha\beta}(q,p_1,p_2)=
(p_1-p_2)_{\mu}g^{\alpha\beta} + 2q^{\beta}g^{\alpha\mu} 
- 2q^{\alpha}g^{\mu\beta}~, 
\label{SMV}
\ee
where the relations $p_{1}^{\alpha}J_{\alpha}^{W^{-}}=0$, 
$q_{\mu}\epsilon^{\mu}_{\gamma}(q)=0$, and 
$p_{2\beta}\epsilon^{\beta}_{W^-}(p_2)=0$ have been used.
The anomalous piece $\Gamma^{\mu\alpha\beta}_{an}$
reads 
\footnote{
Since we only consider the $\gamma WW$ vertex and no 
confusion can arise between the corresponding couplings of the $ZWW$ vertex 
we suppress any super(sub)scripts $\gamma$ 
from all form-factors}
\bea
\Gamma^{\mu\alpha\beta}_{an}(q,p_1,p_2)&=&
\frac{\lambda}{M_W^2} \ (p_2-p_1)^{\mu}q^{\alpha}q^{\beta} + 
(\Delta\kappa+\lambda~)
q^{\beta}g^{\alpha\mu} - \Bigg(\Delta\kappa+\frac{1}{\rho}\lambda\Bigg)
q^{\alpha}g^{\mu\beta} \nonumber \\
&& +
\Bigg[\tilde{\kappa}-
\left(\frac{1+\rho}{2\rho}\right)\tilde{\lambda}\Bigg]
\varepsilon^{\mu\alpha\beta\rho}q_{\rho}\nonumber \\
&&-
\frac{\tilde{\lambda}}{2M_W^2}\ \left[ 
(p_2-p_1)^{\mu}\epsilon^{\alpha\beta\rho\sigma}q_{\rho}(p_2-p_1)_{\sigma}
 +s(\rho-1)
\epsilon^{\mu\alpha\beta\rho}(p_2-p_1)_{\rho}\right]
\eea 
where $\rho\equiv M_W^2/s$. The anomalous couplings 
$\Delta\kappa$, $\lambda$, $\tilde{\kappa}$, and 
$\tilde{\lambda}$ parametrize the 
deviations from the Standard Model vertex; 
they are form-factors which must be evaluated at the 
kinematical point relevant for the reaction under consideration, i.e. 
\be
z\equiv z(p_1^2=s,p_2^2=M_W^2,q^2=0)~,
\ee
where $z$ stands for any of the four aforementioned couplings.
The anomalous vertex given above is  
the most general $\gamma WW$ vertex compatible with Lorentz and $U(1)$ 
gauge invariance; it has been derived from the interaction Lagrangian 
\bea 
\frac{1}{e}{\cal L}_{\gamma WW} &=& 
 \left(ig_1 W^{\dagger}_{\lambda\nu}W^{\lambda}A^{\nu} + h.c.\right) 
 + i \kappa W^{\dagger}_{\lambda}W_{\nu}F^{\lambda\nu} 
 + i \frac{\lambda}{M_W^2}
  W^{\dagger}_{\rho\lambda}W^{\lambda}_{\nu}F^{\nu\rho} \nonumber\\
&&+\frac{g_4}{M_W^2}W^{\dagger}_{\lambda}W_{\nu}
\left( \partial^{\lambda}\partial_{\rho}F^{\rho\nu}
+\partial^{\nu}\partial_{\rho}F^{\rho\lambda}\right)
-\left(\frac{g_5}{M_W^2}\epsilon^{\lambda\nu\rho\sigma}
W^{\dagger}_{\lambda}\partial_{\rho}W_{\nu}
\partial^{\tau}F_{\tau\sigma} +h.c.\right)\nonumber\\\
&&+i {\tilde\kappa} W^{\dagger}_{\lambda}W_{\nu}
{\widetilde F}^{\lambda\nu} 
 + i \frac{{\tilde\lambda}}{M_W^2}
  W^{\dagger}_{\rho\lambda}W^{\lambda}_{\nu}{\widetilde F}^{\nu\rho}~,
\label{L}
\eea
where $W^{ \mu}$ is the $W^-$ field, $A^{\mu}$ is the photon field, the 
field strength tensors are all Abelian, and given by 
$W^{\mu\nu}=\partial^{\mu}W^{\nu}-\partial^{\nu}W^{\mu}$,
$F^{\mu\nu}=\partial^{\mu}A^{\nu}-\partial^{\nu}A^{\mu}$, 
${\widetilde F}^{\mu\nu}=
\frac{1}{2}\epsilon^{\mu\nu\rho\sigma}F_{\rho\sigma}$,
and $\Delta\kappa \equiv \kappa - 1 $ .
The reason why 
the form-factors
$g_1$, $g_4$, and $g_5$ do not appear in $\Gamma^{\mu\alpha\beta}_{an}$
is that by gauge-invariance $g_1$ is forced 
to be fixed at the value $g_1(p_1^2,p_2^2,q^2)=1$, whereas $g_4$ and $g_5$  
to be  proportional (in momentum space) to $q^2$, and thus to 
vanish in our case of an on-shell photon. 
\cite{Hagi},\cite{BZ88},\cite{BB90},\cite{DZ},\cite{SW}.

It is straightforward to verify that 
 the full vertex $\Gamma^{\mu\alpha\beta}$
satisfies the elementary Ward identity 
\be
q_{\mu}\Gamma^{\mu\alpha\beta}(q,p_1,p_2) = 
(p_2^2g^{\alpha\beta}-p_2^{\alpha}p_2^{\beta})-
(p_1^2g^{\alpha\beta}-p_1^{\alpha}p_1^{\beta})~,
\label{WIvertex}
\ee
a fact which guarantees the electromagnetic gauge-invariance 
of the entire amplitude, i.e.
$q^{\mu}T_{\mu\beta}=0$.

The squared unpolarized amplitude for the process is given by 
\be\label{TT}
\sum_{s_{\bar{u}},s_{d}}
\sum_{\lambda_{\gamma},\lambda_{W}}
|\langle d \bar{u}|T| W^{-}\gamma \rangle|^{2}
= \sum_{s_{\bar{u}},s_{d}} T_{\mu\beta} P^{\mu\mu'}(q) Q^{\beta\beta'}(p_2) 
T_{\mu\beta}^{\dagger}~,
\ee
where the sum on $s_{\bar{u}}$, $s_{d}$
runs over all possible helicities of the incoming quarks,
$Q_{\beta\beta'}$ is
the polarization sum of the $W$, given by 
\be
Q^{\beta\beta'}(p_2)=
-g^{\beta\beta'}+\frac{p_{2}^{\beta} p_{2}^{\beta'}}{M_W^2}~,
\ee
and $P^{\mu\mu'}(q)$  the polarization sum of the photon, given by 
\begin{equation}
P^{\mu\mu'}(q)\ =\ -g^{\mu\mu'}+ \frac{\eta^{\mu}q^{\mu'}
+\eta^{\mu'}q^{\mu} }{\eta q} - 
\eta^2 \frac{q^{\mu}q^{\mu'}}{{(\eta q)}^2}\, ,
\label{PhotPol}
\end{equation}
where $\eta_{\mu}$ is an arbitrary four-vector, 
analogous to a gauge-fixing parameter. 
By virtue of the Ward identity of Eq.\ (\ref{WIvertex}) satisfied by 
$\Gamma^{\mu\alpha\beta}$ 
all dependence on $\eta_{\mu}$ disappears from the 
righ-hand side (RHS) of  Eq.\ (\ref{TT}).

The unpolarized color averaged differential cross-section in the 
center-of-mass
frame is given by
\be 
\Bigg(\frac{d\sigma}{dx}\Bigg)= \frac{1}{3}\Bigg(\frac{1-\rho}{32\pi s }\Bigg)
\sum_{s_{\bar{u}},s_{d}}\sum_{\lambda_{\gamma},\lambda_{W}}
|\langle d \bar{u}|T| W^{-}\gamma \rangle|^{2}~. 
\ee
where $x\equiv \cos\theta$, and $\theta$ is the scattering angle 
of the photon relative to the incoming anti-quark
in the center-of-mass frame of the two partons, or equivalently, the 
produced $W$ and $\gamma$.
The differential cross-section is the sum of two pieces, 
\be
\Bigg(\frac{d\sigma}{dx}\Bigg) = 
   \Bigg(\frac{d\sigma^{0}}{dx}\Bigg)+\Bigg(\frac{d\sigma^{an}}{dx}\Bigg)~.
\label{CS}
\ee
The standard contribution
$(d\sigma^{0}/dx)$
corresponds to the case where all anomalous couplings 
are set equal to zero, and has been studied
extensively in the literature 
\cite{MSS},\cite{Robi},\cite{Hagi},\cite{BZ88},\cite{BB90}.
In particular 
it exhibits a radiation zero at $x=-1/3$:   
\be
\frac{d\sigma^{0}}{dx}=\bigg(\frac{1}{3}\bigg)
\frac{e^2g^2}{128\pi}\frac{1}{s(1-\rho)}
\frac{(x+1/3)^2[(1+\rho)^2+(1-\rho)^2x^2}{1-x^2}.
\ee
The anomalous contribution $(d\sigma^{an}/dx)$ is given by
\be
\Bigg(\frac{d\sigma^{an}}{dx}\Bigg) = 
C(s) \Bigg [\sigma_1(s)P_1(x)+
\left ( \frac{1}{2\rho}\right)
\sigma_2(s)P_2(x)+ \left (\frac{1}{4\rho^2}\right)
\sigma_3(s)P_3(x)\Bigg]~,
\label{dsigmaan}
\ee
\be 
C(s)=\bigg(\frac{1}{3}\bigg)\frac{e^2g^2}{256\pi } \frac{(1-\rho)}{s}~,
\ee
where
\bea
P_1(x) & =& x+3x^2  ~,\nonumber\\
P_2(x) & =& 1~,\nonumber\\ 
P_3(x) & =& 1-x^2~,
\label{Pol}
\eea
and
\bea
\sigma_1(s)& =& -\frac{2}{3} \Delta\kappa~, 
\nonumber\\
\sigma_2(s) &=& (\Delta\kappa+\lambda)^2 + 
(\tilde{\kappa}+ \tilde{\lambda})^2~, 
\nonumber\\
\sigma_3(s) &=& 2(\lambda^2+\tilde{\lambda}^2)-\rho[(\Delta\kappa-\lambda)^2
+(\tilde{\kappa}-\tilde{\lambda})^2] 
+ 2\rho^2 (\Delta\kappa^2 +\tilde{\kappa}^2)~.    
\label{sigma3}
\eea
The expression given in Eq.(\ref{dsigmaan}) applies also to the process 
$u\overline{d}\rightarrow W^+\gamma$, with the only difference
that, in that case, the scattering angle 
is defined
between the incoming quark and the photon. Our result agrees with 
previous calculations which considered only $\Delta\kappa$ \cite{MSS}, 
or both the $\Delta\kappa$ and $\lambda$ couplings \cite{Robi}. 

It is well known that,
due
to the presence of the terms $\rho^{-1}$ and $\rho^{-2}$,
the cross-section of Eq.(\ref{dsigmaan}) 
would grossly violate unitarity 
at high energies, $s\rightarrow\infty$, $\rho\rightarrow 0$, if 
the anomalous couplings were considered to be constants, independent of 
$s$. Therefore, in analogy to nuclear form-factors, one 
traditionally assumes an $s$-dependence of the form \cite{BZ88}  
 \be
 z(s,M_W^2,0)=\frac{z_0}{\left(1+s/\Lambda^2\right)^n} ~.
\label{nuc} 
\ee  
 From the explicit expression of the cross-section we see that the choices
 $ n=1/2$ for $\Delta\kappa$, $\tilde{\kappa}$, and 
 $n=1$ for $\lambda$, $\tilde{\lambda}$ 
 would suffice in order for the square brackets in Eq.(\ref{dsigmaan}) 
 to approach  
 a constant as $\rho\rightarrow 0$; then the cross section 
 would decrease like $1/s$ due to the flux factor in  $C(s)$.
 The scale $\Lambda$ is assumed to 
 be of the same order of magnitude as the scale characterising the 
new physics responsible for the anomalous couplings.
 
We note that the coupling $\lambda$ contributes only quadratically 
in the cross section. 
We have no simple explanation of this fact,
 which seems to be characteristic to the specific process.
Of course the $C$, $P$ violating couplings $\tilde{\kappa},\tilde{\lambda}$
can only contribute quadratically. 
Thus, if the anomalous couplings are small enough 
($\leq 10^{-3}$) so that for certain  center of mass energies 
all quadratic terms could be neglected as unobservable, then 
any deviation 
would be a direct measurement of $\Delta\kappa$. 
This is however not the case for typical  Tevatron and LHC energies;
therefore, quadratic terms must be kept in general, and 
are typically of the same order, or even larger than the linear one. 

It is interesting to notice that 
the term linear in $\Delta\kappa$ 
does {\it not} distort the appearance of the
radiation zero at $x=-1/3$, for arbitrary values of $\Delta\kappa$. 
The radiation zero is washed out only due to 
the quadratic anomalous terms, which
reduce it to a dip.
Since the quadratic coefficients of $\lambda$ 
 are much larger than those of $\Delta\kappa$, due to the 
pre-factor $(1/4\rho^2)$ in front of 
 $\sigma_3$ in Eq.(\ref{dsigmaan}),   
  the cross section is extremely sensitive to non standard $\lambda$ values
in the neighbourhood of the radiation zero. 
 
We also notice that the quadratic terms are completely symmetric 
under  
$(\Delta\kappa,\lambda)\leftrightarrow (\tilde{\kappa},\tilde{\lambda})$. 
Furthermore the coefficient in front 
of $\Delta\kappa$ is independent of the energy. 
Thus in experimental analyses 
where the fitting is carried out by allowing
one of the anomalous couplings to  
differ from zero at a time, the $\lambda$ and $\tilde{\lambda}$ 
distributions as well as other related bounds would be identical. 
The same would be true
for $\Delta\kappa$ and $\tilde{\kappa}$, if the term linear in  
$\Delta\kappa$ were negligible; this could be the case 
at high energies, 
if $\Delta\kappa$ is not very small ($\Delta\kappa>10^{-2}$).

\setcounter{equation}{0}
\section{Projecting out the anomalous couplings}

In this section we show how one can extract experimental 
values for the quantities 
$\sigma_1$, $\sigma_2$, and $\sigma_3$, defined in 
Eq.(\ref{sigma3}). This will furnish a system of
three independent algebraic equations involving
the four unknown gauge couplings.

To accomplish this, one first 
notices that the polynomials $P_{i}(x)$ constitute a
linearly-independent set; their Wronskian $W(P_i)$ is simply $W(P_i)=2$. 
One may then   
construct
a set of three other polynomials, $\widetilde{P}_i(x)$, which are
orthonormal 
to the $P_i(x)$, i.e. they satisfy 
\be
\int_{-1}^{1}\widetilde{P}_i(x)P_j(x) dx = \delta_{ij} ~.
\ee
These polynomials are
\footnote{Of course this set is not uniquely determined; here
we derive the set with the lowest possible degree in $x$.} : 
\bea
\widetilde{P}_1(x)& =& \frac{3}{2}x~,    \nonumber\\
\widetilde{P}_2(x) &=& - \frac{3}{4}-\frac{9}{2}x +\frac{15}{4}x^2~, 
\label{projpol}  
\nonumber\\
\widetilde{P}_3(x)& =& \frac{15}{8}+\frac{9}{2}x -\frac{45}{8}x^2~. 
\eea
It is important to observe that the polynomials $P_i (x)$ and 
$\tilde{P_i}(x)$
are {\it independent} of the (sub)-process energy $s$.
This is to be contrasted to the 
equivalent set of polynomials obtained in the
context of the process $e^{+}e^{-} \to W^{+}W^{-}$ \cite{sumrules},
which depend explicitly on the (fixed) $s$. 
As we will
explain  
in section 5 the $s$-independence of the polynomials 
$\tilde{P_i}(x)$
is crucial for the applicability of
the proposed method to the realistic case of $p\bar{p}$ and 
$pp$ 
scattering, where the structure functions must be included.

By means of the polynomials $\tilde{P}_i(x)$ 
one may then invert Eq.(\ref{sigma3}), and 
project out the individual $\sigma_i$ as follows:
\bea
\sigma_1 (s) &=& C^{-1}(s) \int_{-1}^{1} dx   
\Bigg(\frac{d\sigma^{an}}{dx}\Bigg)\widetilde{P}_1(x) ~,\nonumber\\
\sigma_2 (s) &=& 4\rho^2 C^{-1}(s) \int_{-1}^{1} dx   
\Bigg(\frac{d\sigma^{an}}{dx}\Bigg)\widetilde{P}_2(x) ~,\nonumber\\
\sigma_3 (s) &=& 4\rho^3 C^{-1}(s) \int_{-1}^{1} dx   
\Bigg(\frac{d\sigma^{an}}{dx}\Bigg)\widetilde{P}_3(x) ~,
\label{SIG}
\eea
In order to extract the experimental values 
$\sigma_i^{exp}$
for the observables
$\sigma_i$ we simply substitute in the left-hand side of
Eq.(\ref{SIG})
the experimental 
value $(d\sigma^{an}_{exp}/dx)$ given by
\be
\Bigg(\frac{d\sigma^{an}_{exp}}{dx}\Bigg)=
\Bigg(\frac{d\sigma_{exp}}{dx}\Bigg)- 
\Bigg(\frac{d\sigma^{0}}{dx}\Bigg)~.
\ee

We notice from Eq.(\ref{sigma3})
that $\sigma_1^{exp}$ would determine {\it directly}
the experimental value for $\Delta\kappa$, {\it without} 
any assumptions on the size of the other three couplings.
The remaining two equations involving
$\sigma_2^{exp}$ and 
$\sigma_3^{exp}$ can then be used as sum rules, in order
to impose experimental constraints on the three 
remaining couplings $\lambda$,
$\tilde{\kappa}$, and $\tilde{\lambda}$.

In the case when experimental cuts 
restrict the angular region from $[a,b]$ 
instead of [-1,1] appropriate orthonormal polynomials $\widetilde{P}_i(x)$
can still be 
easily constructed, by requiring
\be
\int_{a}^{b}\widetilde{P}_i(x)P_j(x) dx = \delta_{ij} ~.
\ee
Their closed form reads
\be
\widetilde{P}_i(x) = c_{i0}+ c_{i1}x+ c_{i2}x^2 ~, i=1,2,3
\ee
with 
\bea
c_{10} &=& 36D(a^3+4a^2b+4ab^2+b^3)~, \nonumber\\
c_{11} &=& -48D(4a^2+7ab+4b^2)~, \nonumber\\
c_{12} &=& 180D(a+b)~,\nonumber\\
c_{20} &=& -3D\Bigg[3a^4+12a^3(3+b)+(10+36b+3b^2)(b^2+4ab)
+2a^2(5+72b+15b^2)\Bigg]~,\nonumber\\
c_{21} &=& 36D\Bigg[a^3+4a^2(4+b)+b(5+16b+b^2)+a(5+28b+4b^2)\Bigg]~,
\nonumber\\
c_{22} &=& -30D\Bigg[6+a^2+18b+b^2+2a(9+2b)\Bigg]~,\nonumber\\
c_{30}&=& 6D\Bigg[18a^3+(5+18b)(4a+b)b+a^2(5+72b)\Bigg]~,\nonumber\\
c_{31}&=& -36D\Bigg[16a^2+b(5+16b)+a(5+28b)\Bigg]~,\nonumber\\
c_{32}&=& 180D(1+3a+3b)~,
\label{ab}
\eea
where 
\be
D\equiv (a-b)^{-5}~.
\ee
As a check one may verify that the choice $a=-1$ and $b=1$ in 
Eq.(\ref{ab}) reproduces the set given in Eq.(\ref{projpol}).

\setcounter{equation}{0}
\section{Statistical properties of the $\sigma_i$ observables }

In this section we will present an elementary study
of the basic statistical properties of the $\sigma_i$ observables 
introduced in the previous section. In particular we will 
compute their correlations, using simple assumptions about
the distribution of the anomalous couplings.
 For convenience, in this section we introduce 
the following uniform notation:
\be
z_1 \equiv \Delta\kappa, 
~~z_2 \equiv \lambda, 
~~z_3 \equiv \tilde{\kappa},~~
z_4 \equiv \tilde{\lambda}~. 
\ee

To study the correlations of the $\sigma_i$ observables 
we will that assume each of the couplings $z_i$
obeys independently a normal (Gaussian) probability distribution,
with mean $\mu_i$ and variance $ \delta_i^2$, i.e. 
\be
p_{i}(z_i, \mu_i, \delta_i^2) = \frac{1}{\delta_i (2\pi)^{\frac{1}{2}}}
\exp \Bigg [ -\frac{(z_i- \mu_i)^2}{2\delta_i^2}\Bigg ]~.  
\ee
Then, the expectation value
$\langle \sigma_i\rangle$ of the observable $\sigma_i$ 
is given by
\be
\langle \sigma_i\rangle = \prod\limits_{j=1}^{4}
\int_{-\infty}^{+\infty} 
[dz_{j}] p_j \sigma_i~,
\ee
the corresponding covariance matrix by
\be
V_{ij}= \langle \sigma_i \sigma_j \rangle -
\langle \sigma_i \rangle \langle \sigma_j \rangle ~,
\ee
and the correlations $r_{ij}$  by
\be
r_{ij} = \frac{V_{ij}}{V_{ii}^{1/2}V_{jj}^{1/2}}~.
\ee

We will next assume that the Gaussian distribution is
peaked around the Standard Model  
values of the couplings, i.e. $\mu_i=0$, 
and will use the elementary results
\be
\int_{-\infty}^{+\infty}[dz_{i}]
 p_{i}^{(0)} z_i =0,~~
\int_{-\infty}^{+\infty}[dz_{i}]z_i^2
 p_{i}^{(0)} =\delta_i^2,~~
\int_{-\infty}^{+\infty}[dz_{i}] z_i^4
 p_{i}^{(0)} =\frac{3}{4}\delta_i^4 ~,
\ee
where $p_{i}^{(0)}\equiv p_{i}(z_i,0, \delta_i^2)$. 
After a straightforward calculation
we obtain the following expressions for the various $V_{ij}$:
\bea
V_{11} &=& \frac{4}{9} \delta_1^2 ~,\nonumber\\
V_{12} &=& 0 ~,\nonumber\\
V_{13} &=& 0 ~,\nonumber\\       
V_{22} &=& 4\delta_1^2\delta_2^2
          + \frac{1}{2} \delta_1^4
          + \frac{1}{2}\delta_2^4
          + 4\delta_3^2\delta_4^2
          + \frac{1}{2}\delta_3^4
          + \frac{1}{2}\delta_4^4 ~,
\nonumber\\
V_{33} &=& -2\rho (\delta_2^4 + \delta_4^4 )
       +  \frac{1}{2}\rho^2 (
           8\delta_1^2 \delta_2^2
          + \delta_1^4
          + \delta_2^4
          + 8\delta_3^2 \delta_4^2
          + \delta_3^4
          + \delta_4^4
          )\nonumber\\
&&      -2\rho^3 (\delta_1^4 + \delta_3^4 )
       + 2\rho^4 (\delta_1^4 + \delta_3^4)
       + 2 (\delta_2^4 + \delta_4^4) ~,
\nonumber\\
 V_{23} &=&
         \frac{1}{2} \rho (
          8 \delta_1^2 \delta_2^2
          - \delta_1^4
          - \delta_2^4
          + 8 \delta_3^2 \delta_4^2
          - \delta_3^4
          - \delta_4^4
          )\nonumber\\
&&       + \rho^2 (
           \delta_1^4 + \delta_3^4)
       + \delta_2^4
          + \delta_4^4 ~.
\eea    

Evidently, the only non-zero correlation is $r_{23}$. By varying the 
value of the corresponding variances $\delta_i$ one can 
study the limits where some of the 
couplings are excluded on the grounds
of certain discrete symmetries. For example, by choosing 
$\delta_3$ and $\delta_4$ very small compared to $\delta_1$ and 
$\delta_2$
we approach the limit where the $C$ and $P$ symmetries are individually
respected by the anomalous couplings. The other interesting parameter
to vary is of course 
the center-of-mass energy $s$. In the following table
we display  some characteristic cases:

\begin{center}
\begin{tabular}{|c|l|l|l|l|l|}\hline
$\sqrt s$ (GeV) & 100 & 200& 300 & 500 & 1000 \\ \hline\hline  
$r_{23}$ & 0.93& 0.50 & 0.40 & 0.35 & 0.32 \\ \hline\hline
\end{tabular}
\end{center}
\noindent ~~~~~~~~~~~~~~~~~~~~~~~~~~~~~~~~~~~~~~~~~~~~~~
$\delta_1^2=\delta_2^2 \ge \delta_3^2=\delta_4^2$
\vspace*{0.2cm}

\begin{center}
\begin{tabular}{|c|l|l|l|l|l|}\hline
$\sqrt s$(GeV) & 100 & 200& 300 & 500 & 1000 \\ \hline\hline  
$r_{23}$ & 0.92& 0.44 & 0.32 & 0.25 & 0.23 \\ \hline\hline
\end{tabular}
\end{center}
\noindent  ~~~~~~~~~~~~~~~~~~~~~~~~~~~~~~~~~~~~~~~~~~~~~~
$\delta_1^2=2\delta_2^2 = 10\delta_3^2=10\delta_4^2$
\vspace*{0.2cm}

\begin{center}
\begin{tabular}{|c|l|l|l|l|l|}\hline
$\sqrt s$(GeV) & 100 & 200& 300 & 500 & 1000 \\ \hline\hline  
$r_{23}$ & 0.91& 0.40 & 0.27 & 0.21 & 0.18 \\ \hline\hline
\end{tabular}
\end{center}
\noindent  ~~~~~~~~~~~~~~~~~~~~~~~~~~~~~~~~~~~~~~~~~~~~~~
$\delta_1^2=3\delta_2^2=50\delta_3^2=50\delta_4^2$
\vspace*{0.2cm}

\noindent {\bf Table 1} : The correlation $r_{23}$ 
as a function of
the center-of-mass energy $s$,
for 
various choices of the parameters $\delta_i$.
\vspace*{0.5cm}

We see that, with the exception of relatively low energies, the 
values of $r_{23}$ are rather reasonable, and they
tend to decrease as the sub-process energy $s$ increases.

\setcounter{equation}{0}
\section{Inclusion of the structure functions}

The analysis presented so far is valid at the partonic level,
and the quarks appearing in the initial state were assumed to be 
in their center-of-mass frame.
 In reality
the initial states are protons and anti-protons, a fact which introduces
two additional complications. First, the final state can be reached 
by different combinations of partons, which are not in their 
 center-of-mass
any more, but carry  momentum fractions 
$x_a$ and $x_b$ of the corresponding parent hadrons. This is taken into 
account by introducing structure functions. 
Second, the partonic center-of-mass frame 
has to be reconstructed from the data on an event by event basis.
 However, not all kinematical information 
on the final state particles is  available, since the 
final $W$ boson decays to a lepton and
an (unobserved) neutrino. 
Although the transverse momentum of the neutrino 
is identified 
with the missing transverse momentum of the event, its longitudinal momentum 
 can only be 
determined with a twofold ambiguity by constraining the lepton-neutrino pair 
invariant mass to equal the $W$ mass \cite{BZ88},\cite{BB90}. 
Using the fact that, 
at least at the Tevatron,
the $W$ boson
is highly polarized, one can arrive at
the correct choice with a success-rate of 73\% \cite{cdf1}. 

For a $p\bar{p}$ collider the total 
cross-section   
for the process $p\bar{p}\to W^{-}\gamma X$, reads \cite{MSS}
\bea
\frac{d\sigma_{p\bar{p}}(S,x)}{dx}
&&=~~
\sum_{u,d}
\int_0^1 \int_0^1 dx_{a}dx_{b}f_{d/p}(x_a,s)
f_{\bar{u}/\bar{p}}(x_b,s)
\frac{ d\sigma_{d\bar{u}} (s,\cos\theta_{1})}{ d\cos\theta_{1}}
\nonumber\\
&&+~~ 
\sum_{u,d}
\int_0^1 \int_0^1 dx_{a}dx_{b}
f_{\bar{u}/p}(x_a,s)f_{d/\bar{p}}(x_b,s)
\frac{d\sigma_{\bar{u}d}(s,\cos\theta_{2})}{d\cos\theta_{2}}~,
\label{MSS1}
\eea
where $S$ is the square of the $p\bar{p}$  center-of-mass energy, 
$S=(p_{p}+p_{\bar{p}})^2$, 
$x_{a}$ and $x_{b}$ are the fractional longitudinal momenta of the
quarks inside the proton, $0\le x_{a,b}\le 1$,
and 
$s=x_{a}x_{b}S$ is the squared  center-of-mass 
energy of the sub-process. 
In the above equation $u$ denotes a generic $up$-type quark ($u,c,t$), while 
$d$ a generic $down$-type quark ($d,s,b$). 
The double sum runs over all possible 
combinations of $up$ and $down$ quarks ($\bar{u}d,~\bar{u}s,~\bar{c}d,...$) 
which give rise to the desired final state $W^-\gamma$, and for which the 
corresponding structure functions are not negligible. 
$d\sigma_{d\bar{u}}$ denotes the situation where the $d$ quark originates 
from the proton  and the $\bar{u}$ from the anti-proton beam; in that case  
valence-quark structure functions are involved for both quarks.  
$d\sigma_{\bar{u}d}$ denotes the reverse situation; now both structure 
functions involve sea-quarks and are equal. We  define the $W\gamma$ 
center-of-mass
angle $x=\cos\theta_{c.m.}$, to be the angle between the 
photon and the anti-proton beam. Thus, 
the angle involved in  $d\sigma_{d\bar{u}}$
is $\theta_1=\theta_{c.m.}$, while in  $d\sigma_{\bar{u}d}$ 
it is  $\theta_2=\pi-\theta_{c.m.}$, or $\cos\theta_1= x$ and 
$\cos\theta_2=-x$. Clearly, we have that 
$d\sigma_{\bar{u}d}(s,x)=d\sigma_{d\bar{u}}(s,-x)$. 

Notice that if the structure function weight  $f(x_a,s)f(x_b,s)$ 
multiplying each term were the 
same, then all terms linear in $x$ would cancel
in Eq.~(\ref{MSS1}); in such a case the polynomial 
$P_1 (x) \to {\bar{P}}_1(x)=3x^2$,
a fact which would render the set of polynomials ${\bar{P}}_1(x) $,
$P_2(x)$, $P_3(x)$
linearly  
dependent, thus reducing 
the usefulness of the  proposed method.
This is however not the case, since 
the first term on the RHS of
Eq.~(\ref{MSS1}) originates from the valence quarks inside the
proton and anti-proton, whereas  the second term comes from the 
sea-quarks. The sea-quark contribution 
is significantly smaller than that of the valence-quarks,
a fact which is appropriately encoded in the form of the 
corresponding structure functions,  
convoluted with the above elementary processes. 
One could therefore, to a good approximation, 
omit this term. 
If, nonetheless, such a term were to be kept, the necessary procedure 
would be as 
follows: 
Since $d\sigma_{\bar{u}d}(s,x)=d\sigma_{d\bar{u}}(s,-x)$, 
 it follows from Eq.~(\ref{dsigmaan}) and  Eq.~(\ref{Pol})
that
$d\sigma_{\bar{u}d}(s,x)$ is a 
linear combination of $P_2(x)$, $P_3(x)$, and 
$\widehat{P}_1(x) = P_1(-x)=3x^2-x$. 
The next step is to write $\widehat{P}_1(x)$ 
as a linear combination of $P_1$(x), $P_2(x)$, and $P_3(x)$, i.e.
\be
\widehat{P}_1(x) = -P_1(x) +6P_2(x) -6P_3(x)~, 
\label{lin}
\ee
after which
the quantities $\sigma_1$, $\sigma_2$, and  
$\sigma_3$ will be simply modified by small corrections.

Omitting for simplicity such corrections from sea-quarks, 
and keeping only 
the $(\bar{u} d)$ term in the sum on the RHS, Eq.~(\ref{MSS1})
becomes
\be
\frac{d\sigma_{p\bar{p}}(S,x)}{dx}
=
\int_0^1 \int_0^1 dx_{a}dx_{b}f_{d/p}(x_a)
f_{\bar{u}/\bar{p}}(x_b)
\frac{d\sigma_{\bar{u}d}(s,x)}{dx}~,
\ee
and its deviation 
from the canonical value due to the anomalous gauge boson couplings is
given by
\bea
\frac{d\sigma_{p\bar{p}}^{(an)}(S,x)}{dx}
&=&
\int_0^1 \int_0^1 dx_{a}dx_{b}f_{d/p}(x_a)
f_{\bar{u}/\bar{p}}(x_b)
\frac{d\sigma_{\bar{u}d}^{(an)}(s,x)}{dx}
\nonumber\\
&=&
\int_0^1 \int_0^1 dx_{a}dx_{b}f_{d/p}(x_a)
f_{\bar{u}/\bar{p}}(x_b)
\times \nonumber \\ 
&&~C(s) \Bigg [\sigma_1(s)P_1(x)+
\left ( \frac{1}{4\rho^2(s)}\right)
\sigma_2(s)P_2(x)+ 
\left (\frac{1}{4\rho^3(s)}\right)
\sigma_3(s)P_3(x)\Bigg]~.
\label{spp}
\eea

At this point we can repeat the same procedure for projecting out  
the anomalous couplings which has been presented in section 3. 
This is possible
because the polynomials
$P_i(x)$ and the corresponding 
projective polynomials $\tilde{P}_i(x)$ do {\it not} depend
on the sub-process energy $s$.
Writing Eq.(\ref{spp}) in the form   
\be
\frac{d\sigma_{p\bar{p}}(S,x)}{dx} = \sum_{i=1}^{3}\Sigma_i(S) P_i(x)
\ee
with
\bea
\Sigma_1(S) &\equiv & \int_0^1 \int_0^1 dx_{a}dx_{b}f_{d/p}(x_a)
f_{\bar{u}/\bar{p}}(x_b)C(s)\sigma_1(s) \nonumber\\
\Sigma_2(S) &\equiv & \int_0^1 \int_0^1 dx_{a}dx_{b}f_{d/p}(x_a)
f_{\bar{u}/\bar{p}}(x_b)\Bigg(\frac{C(s)}{4\rho^2(s)}\Bigg)
\sigma_2(s) \nonumber\\
\Sigma_3(S) &\equiv & \int_0^1\int_0^1 dx_{a}dx_{b}f_{d/p}(x_a)
f_{\bar{u}/\bar{p}}(x_b)
\Bigg(\frac{C(s)}{4\rho^3(s)}\Bigg)\sigma_3(s)~, 
\eea
we see that 
the quantities $\Sigma_i$  
may be extracted from
the differential cross-section by means of the 
projective polynomials $\tilde{P}_i(x)$.
Their experimental value $\Sigma_i^{exp}(S)$ are
obtained from
\be
\Sigma_i^{exp}(S)= \int_{-1}^{1}
\Bigg[\frac{d\sigma_{p\bar{p}}^{(exp)}(S,x)} {dx}
- \frac{d\sigma_{p\bar{p}}^{(0)}(S,x)} {dx}\Bigg]\tilde{P}_i(x)dx~.
\ee
In order to place bounds on the unknown couplings from
the values of $\Sigma_i^{exp}(S)$ one will have to 
take into account the fact that the couplings depend 
on the sub-process energy $s$, i.e. they are functions
of the integration variables $x_a$ and $x_b$.
For example,
assuming an energy dependence such as the one given in 
Eq.(\ref{nuc}), we have for  $\Delta\kappa$ (with $n=1/2$)
\be
\Sigma_1^{exp}(S) = -\frac{2}{3} \Delta\kappa_0 
 \int_0^1 \int_0^1 dx_{a}dx_{b}f_{d/p}(x_a)
f_{\bar{u}/\bar{p}}(x_b)\frac{C(x_ax_bS)}{\sqrt{1+x_ax_bS/\Lambda^2}}~.
\ee
Thus,
measuring $\Sigma_1$ will determine $\Delta\kappa_0$ as a function 
of the arbitrary scale $\Lambda$. 

\medskip

In the case of a $pp$ collider such as the LHC, 
the process $pp\to W^{-}\gamma X$
proceeds only through sea-quark interactions.
The corresponding differential cross-section reads
\bea
\frac{d\sigma_{pp}(S,\cos\theta_{c.m.})}{d\cos\theta_{c.m.}}
&=&
~~\sum_{u,d}
\int_0^1 \int_0^1 dx_{a}dx_{b}f_{d/p}(x_a)
f_{\bar{u}/p}(x_b)
\frac{ d\sigma_{d\bar{u}} }{ d\cos\theta_1 }
\nonumber\\
&&+ 
\sum_{u,d}
\int_0^1 \int_0^1 dx_{a}dx_{b}
f_{\bar{u}/p}(x_a)f_{d/p}(x_b)
\frac{d\sigma_{\bar{u}d}}{d\cos\theta_2}~~~.
\label{MSS2}
\eea
Both terms in the above sum contain the product of
 a valence-quark and 
a sea-quark distribution. This makes the rates of $pp$ 
 cross sections lower than the relative $p\bar{p}$ ones,
a fact which is compensated by the  higher luminosities
of the $pp$ machines.
Therefore,
one important difference between Eq.~(\ref{MSS1})
and Eq.~(\ref{MSS2}) is that, after setting again
$\cos\theta_1 =x$ and $\cos\theta_2 =-x$, 
the terms linear in $x$ cancel in the latter. Indeed,
since the cross-section is symmetric in 
$x_a\leftrightarrow x_b$ both terms
contribute with equal weight under the integral  
and the cross-section assumes the form   
\bea
\frac{d\sigma_{pp}(S,x)}{dx}
&=&
\sum_{u,d}
\int_0^1 \int_0^1 dx_{a}dx_{b}f_{d/p}(x_a)
f_{\bar{u}/p}(x_b) 
\left[\frac{d\sigma_{d\bar{u}}(s,x)}{dx}
+
\frac{d\sigma_{d\bar{u}}(s,-x)}{dx}\right]
\label{ppMSS2}
\eea 

As mentioned before, 
the absence of the linear $x$ term would render the set of polynomials
$P_i$ linearly-dependent.
To avoid this we propose to follow the method of binning the
events according to their longitudinal momentum,
which was introduced in \cite{MSS}. 
The basic point is to break the symmetricity of the 
$\int_0^1\int_0^1 dx_adx_b$
integration
by imposing an asymmetric constraint on the values of
$x_a$ and $x_b$. Let us for example impose the linear constraint
$x_a-x_b\ge\delta$, with $\delta < 1$. Then the ``binned'' differential
cross-section 
$d\sigma_{pp}^{(b)}/d\cos\theta_{c.m.}$
reads
\bea
\frac{d\sigma_{pp}^{(b)}(S,\cos\theta_{c.m.})}{d\cos\theta_{c.m.}}
&=&
~~\sum_{u,d}
\int_{\delta}^{1}\int_{0}^{1-\delta} dx_{a}dx_{b}f_{d/p}(x_a)
f_{\bar{u}/p}(x_b)
\frac{ d\sigma_{d\bar{u}} }{ d\cos\theta_1 }
\nonumber\\
&&+ 
\sum_{u,d}
\int_{\delta}^{1}\int_{0}^{1-\delta} dx_{a}dx_{b}
f_{\bar{u}/p}(x_a)f_{d/p}(x_b)
\frac{d\sigma_{\bar{u}d}}{d\cos\theta_2}~~~.
\label{MSS3}
\eea
or, equivalently,
\bea
\frac{d\sigma_{pp}^{(b)}(S,x)}{dx} &=&
\sum_{u,d}
\int_{\delta}^{1}\int_{0}^{1-\delta} dx_{a}dx_{b}
\left[f_{d/p}(x_a)
f_{\bar{u}/p}(x_b) 
\frac{d\sigma_{d\bar{u}}(s,x)}{dx}
+
f_{\bar{u}/p}(x_a)f_{d/p}(x_b)
\frac{d\sigma_{d\bar{u}}(s,-x)}{dx}\right]
\nonumber\\
&=& \sum_{i=1}^{3}\Sigma_i^{(b)}(S) P_i(x)~,
\eea
with
\bea
\Sigma_1^{(b)}(S) &\equiv & \int_{\delta}^{1}\int_{0}^{1-\delta}dx_{a}dx_{b}
\left[f_{d/p}(x_a)f_{\bar{u}/\bar{p}}(x_b)
-f_{d/p}(x_b)f_{\bar{u}/\bar{p}}(x_a)\right]
C(s)\sigma_1~, \nonumber\\
\Sigma_2^{(b)}(S) &\equiv & \int_{\delta}^{1}\int_{0}^{1-\delta}dx_{a}dx_{b}
\Bigg[
f_{d/p}(x_a)f_{\bar{u}/\bar{p}}(x_b)\frac{\sigma_2}{4\rho^2}
+ 
f_{d/p}(x_b)f_{\bar{u}/\bar{p}}(x_a)
\bigg(\frac{\sigma_2}{4\rho^2}+6\sigma_1\bigg)\Bigg]C(s)~,
\nonumber\\
\Sigma_3^{(b)}(S) &\equiv &  \int_{\delta}^{1}\int_{0}^{1-\delta}dx_{a}dx_{b}
\Bigg[
f_{d/p}(x_a)f_{\bar{u}/\bar{p}}(x_b)\frac{\sigma_3}{4\rho^3}
+ 
f_{d/p}(x_b)f_{\bar{u}/\bar{p}}(x_a)
\bigg(\frac{\sigma_3}{4\rho^3}-6\sigma_1\bigg)\Bigg]C(s)~,
\nonumber\\
&&{}
\eea
where we have used Eq.~(\ref{lin}). Similarly,
the experimental values for the $\Sigma_i^{(b)}(S)$ will be given by
\be
\Sigma_i^{(b,exp)}(S)= \int_{-1}^{1}
\Bigg[\frac{d\sigma_{pp}^{(b,exp)}(S,x)} {dx}
- \frac{d\sigma_{pp}^{(b,0)}(S,x)} {dx}\Bigg]\tilde{P}_i(x)dx~.
\ee
In practice one ought to verify that
the bounds for the anomalous couplings obtained by 
applying the above procedure do not depend 
heavily on how the binning is carried out, by choosing, for example, 
different values
for $\delta$, or different functional forms for the asymmetric
constraint imposed.

\section{Conclusions}

In    this  paper we  have  presented   a model-independent method for
extracting bounds   on   the  anomalous   $\gamma WW$ couplings   from
$pp$ and $p\bar{p}$
experiments, using the process $d\bar{u}\to W^{-}\gamma$ as a prototype.
 At the partonic  level this method   gives rise to  three
observables, which   depend  explicitly    on the   various  anomalous
couplings through  simple algebraic relations.   These observables may
be extracted from the experimentally measured unpolarized differential
cross-section for this process 
by means of a convolution with appropriately constructed
polynomials.  These  polynomials    are quadratic   functions  of  the
center-of-mass angle  only; most  notably, they do not depend  on  the
center-of-mass energy of  the  (sub)-process. One of  these 
observables is   linearly related  to 
$\Delta\kappa$  only;  therefore, its   measurement  can  furnish  the
experimental value of  this quantity,   without further
assumptions on the  values of the  remaining couplings.  The other two
observable  constitute  a system  of two equations  for  the remaining
three anomalous  couplings;  thus they  can be  used as  sum rules, in
conjunction  with  other possible  observables, physically motivated
constraints, or model-inspired relations.

The  generalization of  the  method to  the realistic   case of hadron
colliders, where the initial particles  are not partons but protons or
anti-protons, presents experimental  and  theoretical  complications,
which, however, can be overcome.  From the experimental point of view,
it is clear  that in the  case of hadron colliders the  center-of-mass
frame for the $W\gamma$ must be reconstructed;  the ability to achieve
this  is of course crucial for   the applicability of proposed method,
given that the latter relies heavily on the use of 
the center-of-mass
scattering angle.  One   way to   accomplish  this seems  to  be   the
following: The produced $W$ is in general closely on shell, and highly
polarized;  using the    first fact,  one  can  impose  the constraint
$M_W^2=(p_l+p_{\nu})^2$ to determine  the longitudinal momentum of the
neutrino   with a   twofold  ambiguity   \cite{BB90},\cite{Hagi}  (the
transverse  one  is  the missing   transverse momentum  of the event),
whereas the second fact guarantees that one can select the correct
solution 73\% of the time  \cite{cdf1}.  Thus, the  $W$ momentum can be
reconstructed, and  from it the center-of-mass  angle of the event may
be deduced.

At the theoretical level 
the method carries over straightforwardly 
from the partonic level to the
case of $p\bar{p}$ colliders, since
the inclusion of the structure functions 
does not interfere with any of the underlying assumptions.
After the inclusion of the structure functions 
one needs to assume a certain functional dependence of the unknown
form-factors on the sub-process energies, over which one 
integrates; this is of course a general limitation in this type
of analysis, and is not
particular to this method. 
On the other hand,
in the case of $pp$ colliders the structure functions conspire 
to eliminate the linear terms in $x$, a fact which invalidates one
of the main assumptions, namely that the polynomials $P_i(x)$ are
linearly independent. This may be circumvented if one
considers the ``binned'' instead of the
usual differential cross-section, which
may be obtained by introducing
an asymmetric constraint among the longitudinal momenta appearing
in the arguments of the structure functions \cite{MSS}.

It would be interesting to study whether the 
analysis presented here carries over to the
process $d\bar{u}\to W^{-}Z$, 
which would probe directly and separately
the possible anomalous couplings appearing 
in the $ZWW$ vertex. Finally, we note that
these results can be easily translated to the  $e^{-}\gamma \to W^- \nu_e$ 
process, which would be of interest for Linear Colliders, if the $e\gamma$ 
option is realised.

\vspace{0.7cm}\noindent {\bf  Acknowledgments.}
The work of JP is supported by a EC grant TMR-ERBFMBICT 972024 
and that of KP by a EC grant FMBICT 961033. 
The authors thank W. J. Stirling and M. Stratmann 
for useful discussions.

\newpage
\centerline {FIGURES}

\begin{center}
\begin{picture}(340,200)(0,0)
\SetWidth{0.8}

\ArrowLine(20,150)(50,150)\Text(25,160)[r]{$d (p_d)$}
\ArrowLine(50,150)(50,110)
\ArrowLine(50,110)(20,110)\Text(25,100)[r]{$\bar{u}(p_u)$}
\Photon(50,150)(80,150){3}{3}\Text(105,160)[r]{$\gamma (q)$}
\Photon(50,110)(80,110){3}{3}\Text(120,100)[r]{$W^{-}(p_2)$}

\ArrowLine(140,150)(170,150)
\ArrowLine(170,150)(170,110)
\ArrowLine(170,110)(140,110)
\Photon(170,150)(183,133){3}{3}
\Photon(188,126.3)(200,110){3}{3}
\Photon(170,110)(200,150){3}{5}
\Text(220,160)[r]{$\gamma (q)$}
\Text(230,95)[r]{$W^{-}(p_2)$}

\ArrowLine(240,150)(270,130)
\ArrowLine(270,130)(240,110)
\Photon(270,130)(300,130){3}{3}
\Photon(300,130)(330,150){3}{3}
\Photon(300,130)(330,110){3}{3}
\Vertex(300,130){5}
\Text(350,160)[r]{$\gamma (q)$}
\Text(370,95)[r]{$W^{-}(p_2)$}
\Text(290,120)[r]{$p_1$}

\end{picture}\\
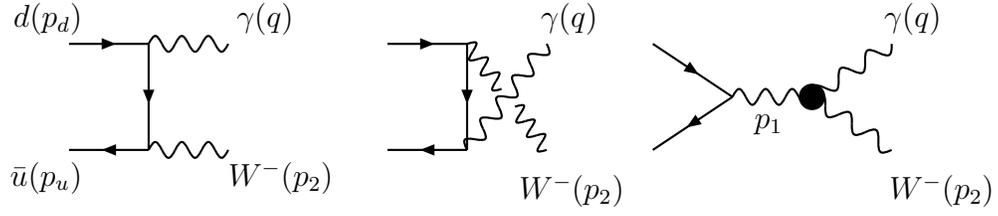
{\bf Fig.\ 1:} Diagrams contributing to the process
$d\bar{u}\to W^{-}\gamma$. The blob at the vertex indicates
the presence of anomalous three-boson couplings. 
\end{center}

\end{document}